\documentclass[aps,prc,floatfix,amsmath,10pt]{revtex4-1}
\usepackage{dcolumn}
\usepackage{amssymb}
\usepackage{mathrsfs}
\usepackage{color,graphicx}
\usepackage[bookmarksopen]{hyperref}
\def  \p    {\pi}

\def  \m    {\mu}

\def  \f    {\frac}

\def  \th   {\theta}

\def  \ra   {\rightarrow}

\def  \veps {\varepsilon}

\def  \del  {\partial}

\def  \bef  {\begin{figure}}
\def  \eef  {\end{figure}}
\def  \be   {\begin{equation}}
\def  \ee   {\end{equation}}
\def  \ba   {\begin{array}}
\def  \ea   {\end{array}}
\def  \bea  {\begin{eqnarray}}
\def  \eea  {\end{eqnarray}}
\def  \beq  {\begin{eqnarray}}
\def  \eeq  {\end{eqnarray}}
\def  \nn   {\nonumber}
\def  \bd   {\begin{displaymath}}
\def  \ed   {\end{displaymath}}
\def  \bse  {\begin{subequations}}
\def  \ese  {\end{subequations}}
\def  \bwt  {\begin{widetext}}
\def  \ewt  {\end{widetext}}

\def  \ba   {{\bf{a_1}}}

\begin{document}
\title{Astrophysical aspects of neutrino dynamics in ultra-degenerate quark gluon plasma}
\author{Souvik Priyam Adhya}
\email{sp.adhya@vecc.gov.in}

\affiliation{Experimental High Energy Physics and Applications Group,
\\
Variable Energy Cyclotron Center,
1/AF Bidhannagar, Kolkata-700 064, INDIA}

\begin{abstract}
The cardinal focus of the present review is to explore the role of neutrinos originating from the ultra-dense core of neutron stars composed of quark gluon plasma in the astrophysical scenario. The collective excitations of the quarks involving the neutrinos through the different kinematical processes have been studied. The cooling of the neutron stars as well as pulsar kicks due to asymmetric neutrino emission have been discussed in detail. Results involving calculation of relevant physical quantities like neutrino mean free path and emissivity have been presented in the framework of non-Fermi liquid behavior as applicable to ultra-degenerate plasma.
\end{abstract}

\maketitle
\section*{Introduction}
Compact stars, in minutes or hours after their birth rapidly cool down to temperatures $(T)$ of the order of $10^9~K$ from upper bounds of $T\succsim 10^{11}~K$. The neutrinos produced in direct reactions at the core are essentially transported to the surface leading to the cooling of the neutron star (NS)\cite{shapiro_book,MSTV90,YLS99,V01,PGW,Sedr07,Huang:2007jw}. It is well known that the matter density in the core of the neutron stars could well exceed up to a few times the nuclear matter saturation density. Thus, in this context, we can expect comparable proportions of up, down and strange quarks \cite{witten84, farhi84} which are quite truly the ground state of QCD at finite baryon density. In addition, at such high densities, the constituents of matter are relativistic which compels one to describe such matter in the framework of non- Fermi liquid (NFL) behavior.
However, the magnetic interaction is suppressed in powers of $(v/c)^2$ for non-relativistic system. In fact, due to the presence of the magnetic contribution, the long range transverse interactions in de-confined degenerate quark matter give rise to the NFL behavior manifested in the appearance of anomalous contribution to the low temperature limit of the quark self energy. The breakdown of the Fermi liquid (FL) picture has drawn substantial theoretical interest primarily due to the detection of the NFL behavior in normal state of superconductors \cite{Gan:1993,Chakravarty:1995,Reizer:1989,Varma:1989} and in the systems of strongly correlated electrons \cite{Polchinski:1992ed,Polchinski:1994ii,Nayak:1994ng}. In a contemporary system such as the quark gluon plasma, at the limit of low temperature, the chromo-magnetostatic fields are un-screened unlike the high temperature case. Thus, we find a logarithmic singularity in the inverse group velocity. This singularity is removed by taking into account Hard dense loop (HDL) approximations to the propagators \cite{manuel00, Bellac97}.
 This recently discovered phenomena of NFL behavior relates itself to the modified quark dispersion relation. Therefore, for excitations close to the Fermi surface, NFL phenomena finds important application in this realm of phase space. It has been shown that the NFL corrections to the quark self-energy enhance the neutrino emissivity of un-gapped quark matter which may exist at the core of neutron stars\cite{schafer04}. Like emissivity, in dense quark matter, the neutrino mean free path(MFP) also receives significant NFL corrections as has been demonstrated in
\cite{pal11}. 
In many calculations, such corrections appear in the modification of the phase space giving rise to appreciable contributions to the well known FL terms. 
Knowledge of the neutrino emissivity showing NFL behavior from quark cores from the neutrino Mean free path (MFP) in compact stars is an essential pre-requisite for a comprehensive understanding of a host of interesting phenomena such as the cooling of the star. 
In view of these contemporary investigations, we here plan to present the
neutrino MFP and corresponding emissivity in normal degenerate quark matter beyond leading logarithmic order and compare it with the leading order (LO) results. Here, LO refers to the anomalous logarithmic term $T \mathrm{log}(1/T)$ that occurs as the first term in the non-Fermi liquid contribution to the fermion self energy. The quantities such as MFP and emissivity of neutrinos calculated with this term is called the LO corrections.  Next to leading order (NLO) terms include all  other terms beyond the LO that contain the fractional powers of $T$ and up to the $(T^3){\rm log}(1/T)$. Quantities calculated with this correction are labeled as NLO corrections \cite{rebhan05}. Knowing the specific heat of dense quark
matter up to the order concerned, we investigate the cooling behavior of the neutron star with dense quark core.
The important theoretical input required for the study of the emissivity and cooling is the neutrino MFP from such quark cores.
 In all these cases, the correction at the LO has been seen to involve $T\ln(1/T)$ term which has been dubbed as anomalous corrections in many of the literatures \cite{holstein73,rebhan05, hieselberg93}. In ref.\cite{Bellac97,manuel00}, it has been shown that the fermion damping rate and energy loss receives significant NFL correction from the LO terms.  Similar behavior has also been reported while investigating quantities like drag , diffusion 
coefficients and thermal relaxation time of electrons for relativistic degenerate plasmas \cite{sarkar10,sarkar11,sarkar13}.
In fact, the specific heat of the quark matter is also modified due to such corrections beyond the FL order \cite{ipp04}. In this review, we will also present the specific heat of degenerate quark matter which receives significant correction due to the presence of the high external magnetic field while including the NFL corrections. Such high magnetic fields are present at the core of the NS. Thus, the inclusion of the magnetic field in the specific heat in the case of interacting plasma and its effect on the kick velocity of the NS are studied in detail.
The  mechanism for the generation of such phenomenal kicks is related to the polarization of the electrons leading to neutrino emission in a preferred direction. The magnetic field strength required has to be equal to or greater than  $B_{cric}=m_{i}^{2} c^{3}/(q_{i}\hbar)$, where $m_i$ and $q_i$ are the mass and charge of the electron. This leads to the formation of the neutrino and anti-neutrino emission cones which give rise to polarized neutrino emission 
opposite to the direction of the magnetic field producing the recoil velocity. 
We have used the electron polarization for different conditions of magnetic field and kick velocities which has been studied recently by Sagert et. al.\cite{sagert08, sagertarxiv1}; where the dependence of the kick velocity on the quark phase temperature and radius of the star with varying quark chemical potentials has also been investigated. 
In fact, experimental efforts like ROSAT
, CHANDRA and XMM have done a lot of measurements to understand
the properties of the neutron stars \cite{Yakovlev:2004iq, finley92}. Thus, at initial times, 
dominant cooling mechanism is by neutrinos only and at later times, by thermal emission.
In addition, if color superconductivity effects are taken into account, the neutrino emissivity and specific heat are then
suppressed exponentially by a factor $exp(-\Delta/T)$ where $\Delta$ is the quasi-particle gap in 
the Color Flavor Locked (CFL) phase. However, the presence of the color superconductivity might lower the
pulsar acceleration \cite{sagert08, Anglani:2006br}.

This review is based upon the recent works \cite{adhya12, adhya14,Adhya:2015uaa, pal11, schafer04, sagertarxiv1, sagert08} where the authors have explored various aspects of the neutrino dynamics in reactions involving ultra-degenerate quark matter. In this review, we discuss the  mean free path of the neutrinos and emissivity of neutrinos from neutron star in Section A and B respectively. In addition, we will present results to see how the cooling is affected when one considers the anomalous corrections into the calculable physical quantities in Section B. In the succeeding section C, we will extend this formalism and introduce the effect of the external magnetic field to calculate the specific heat capacity of degenerate quark matter. This quantity directly influences the kick velocity of pulsars (i.e. rapidly rotating neutron stars) which arises due to asymmetric neutrino emission.
Finally, we will present the results and conclude in Section D.

\subsection{Neutrino mean free path}
\subsubsection*{The case of degenerate neutrinos}
The dominant contribution to the mean free path as well as emission of neutrinos is given by the quark analog of $\beta$ decay and the electron capture reaction producing neutrinos \cite{Iwamoto:1982zz, Lattimer:1991ib, Tubbs:1975jx, Lamb:1976ac}.
The degenerate neutrinos refer to the case where the neutrino chemical potential ($\mu_{\nu}$) is much larger than the temperature.
\beq
d+\nu_{e}\rightarrow u+e^-
\label{dir}
\eeq
\beq
u+e^-\rightarrow d+\nu_{e}.
\label{inv}
\eeq
In addition, the quark-neutrino scattering process is given by,
\beq
q_{i}+\nu_e({\overline \nu_e})\ra q_{i}+\nu_e({\overline \nu_e})
\eeq

  The corresponding mean free paths are denoted by $l^{abs}_{mean}$ and $l^{scatt}_{mean}$ which are subsequently combined to obtain the total mean free path of the neutrinos,
 \bea
 \f{1}{l^{total}_{mean}}=\f{1}{l^{abs}_{mean}}+\f{1}{l^{scatt}_{mean}}
 \eea

The  MFP of the neutrinos is related to the total interaction rate due 
to neutrino emission  which is averaged over the initial quark spins and eventually summed over 
the final state phase space and spins. It is given by \cite{Iwamoto:1982zz},
\bea
\label{mfp01}
\frac{1}{l_{mean}^{abs}(E_{\nu},T)}=&&\frac{g^{\prime}}{2E_{\nu}}\int\frac{d^3p_d}{
(2\p)^3}
\frac{1}{2E_d}\int\frac{d^3p_u}{(2\p)^3}
\frac{1}{2E_u}\int\frac{d^3p_e}{
(2\p)^3 }
\frac{1}{2E_e}
(2\pi)^4\delta^4(P_d +P_{\nu}-P_u
-P_e)\nn\\
&&\times|M|^2 \{n(p_d)[1-n(p_u)][1-n(p_e)]
+n(p_u)n(p_e)[1-n(p_d)]\},
\eea
where, $g^{\prime}$ is the spin and color degeneracy, 
taken as $6$ and $E$, $p$ and $n_{p}$ are the energy, momentum and distribution function for the corresponding particle.
The squared invariant amplitude $|M|^2$ 
 is given by
$|M|^2=64G^2\cos^2\th_c(P_d\cdot P_\nu)(P_u\cdot P_e)$. Here $G\simeq1.435\times10^{-49} erg-cm^{3}$ is the weak coupling constant. The interaction involving strange 
quark is Cabibbo suppressed due to which the neutrino emission from strange quark matter can be neglected. 
We here consider the case of degenerate neutrinos 
{\em i.e.} when $\mu_{\nu}\gg T$. So in
this case both the direct Eq.(\ref{dir}) and inverse 
Eq.(\ref{inv})
processes are relevant.
Therefore, the $\beta$ equilibrium condition becomes $\mu_d+\mu_{\nu}=\mu_u+\mu_e$. 
Now, the momentum integration, $d^{3}p_d$ and $d^{3}p_u$ can be calculated as,
\bea
d^{3}p_d=2\pi \f{p_{f}(d)}{p_{f}(\nu)}pdp\f{dp_{d}}{d\omega}d\omega;
d^{3}p_u=2\pi \f{p_{f}(u)p_{f}(e)}{p}dE_{e}\f{dp_{u}}{d\omega}d\omega
\eea
where we define $p\equiv|p_{d}+p_{\nu}|=|p_{u}+p_{e}|$. $dp(\omega)/d\omega$ can be evaluated from the modified dispersion relation as follows \cite{tatsumi09},
\bea
\f{dp(\omega)}{d\omega}= \Big(1-\f{\del Re\Sigma_{+}(\omega)}{\del \omega}\Big)\f{E_{p(\omega)}}{p(\omega)}
\label{dpdo}
\eea
We use the explicit form of the real part of the quark self energy in the ultra-degenerate limit as found in Refs.(\cite{adhya12,rebhan05,ipp04}) which reads as,
\begin{eqnarray}
\Sigma(\omega)&=&-g^2C_Fm\,
 \Big\{{\varepsilon\over12\pi^2m}\Big[\log\Big({4\sqrt{2}m\over\pi \varepsilon}\Big)+1\Big]+{i \varepsilon\over24\pi m}
  \,+{2^{1/3}\sqrt{3}\over45\pi^{7/3}}\left({\varepsilon\over m}\right)^{5/3}(\mathrm{sgn}(\varepsilon)-\sqrt{3}i)\qquad\nn\\
  &&+ {i\over64 \sqrt{2}}\left({\varepsilon\over m}\right)^2
  -20{2^{2/3}\sqrt{3}\over189\pi^{11/3}}\left({\varepsilon\over m}\right)^{7/3}(\mathrm{sgn}(\varepsilon)+\sqrt{3}i)\qquad\nn\\
&&-{6144-256\pi^2+36\pi^4-9\pi^6\over864\pi^6}\Big({\varepsilon\over m} 
  \Big)^3 \Big[\log\left({{0.928}\,m\over \varepsilon}\right) 
-{i\pi\mathrm{sgn}(\varepsilon)\over 2}  \Big]
  +\mathcal{O}\Big(\left({\varepsilon\over m}\right)^{11/3}\Big) \Big\},
\end{eqnarray}
where $\varepsilon = (\omega-\mu)\sim T$ where NFL effects dominate.
Neglecting the quark-quark interactions, the leading order result is obtained as \cite{pal11, adhya12},
\bea
\frac{1}{l_{mean}^{abs,D}}\Big|_{LO}&&\simeq\frac{2}{3\pi^{5}}G_{F}^{2}C_{F}\cos^{2}\th_{c}\frac{\mu_{e}^{3}}{\mu_{\nu}^{2}}\Big[1+\frac{1}{2}\Big(\frac{\mu_{e}}{\mu}
\Big)
+\frac{1}{10}\Big(\frac{\mu_{e}}{\mu}\Big)^{2}\Big]\nn\\
&&\times[(E_{\nu}-\mu_{\nu})^{2}+\pi^{2} T^{2}](g\mu)^{2}\text{log}\Big(\frac{4g\mu}{\pi^{2}T}\Big).
\eea
The NLO result is evaluated as \cite{adhya12, adhyaconf12,Adhya:2015uaa},
\bea
\frac{1}{l_{mean}^{abs,D}}\Big|_{NLO}&\simeq&\frac{8}{\pi^{3}}G_{F}^{2}C_{F}\cos^{2}\th_{c}\frac{\mu_{e}^{3}}{\mu_{\nu}^{2}}\Big[1+\frac{1}{2}\Big(\frac{\mu_{e}}{\mu}
\Big)
+\frac{1}{10}\Big(\frac{\mu_{e}}{\mu}\Big)^{2}\Big]
[(E_{\nu}-\mu_{\nu})^{2}+\pi^{2} T^{2}]\nn\\
&\times&\Big[a_{1}T^{2/3}(g\mu)^{4/3}+a_{2}T^{4/3}(g\mu)^{2/3}+a_{3}\Big\{1-3 \text{log}\Big(\frac{0.209g\mu}{T}\Big)\Big\}T^{2}\Big]
\eea
where the constants $a_1= 0.015, a_2=-0.075$ and $a_3=-0.036$. For the evaluation of the constants please refer to \cite{adhya12}.
To arrive at the Fermi-liquid result, one can use the free dispersion relation to arrive at \cite{Iwamoto:1982zz},
\bea
\label{mfp_cond1}
\frac{1}{l_{mean}^{abs,D}}\Big|_{FL}=\frac{4}{\pi^{3}}G_{F}^{2}\cos^{2}\th_{c}
\frac{\mu^{2}
\mu_{e}^{3}}{\mu_{\nu}^{2}}\Big[1+\frac{1}{2}\Big(\frac{\mu_{e}}{\mu}
\Big)
+\frac{1}{10}\Big(\frac{\mu_{e}}{\mu}\Big)^{2}\Big]
[(E_{\nu}-\mu_{\nu})^{2}+\pi^{2} T^{2}].
\eea 
 Quarks and electrons are considered to be massless in this framework,
the chemical equilibrium condition gives $p_f(u)+p_f(e)=p_f(d)+p_f(\nu)$,
which we use to derive Eq.(\ref{mfp_cond1}). In addition, we have assumed that $\mu_d \sim \mu_u =\mu$.
The MFP for the quark-neutrino scattering process,
for each quark component of flavor $i (=u~{\rm or}~d)$ is calculated which will contribute to the total neutrino flux.
Proceeding in a similar way, we obtain \cite{Iwamoto:1982zz, Lattimer:1991ib, Tubbs:1975jx, Lamb:1976ac,pal11, adhya12, adhyaconf12},
\bea
\frac{1}{l_{mean}^{scatt,D}}\Big|_{FL}=\f{3}{4\pi}n_{q_i}G_{F}^{2}
\times[(E_{\nu}
-\mu_ { \nu } )^ { 2 } +\pi^{2} T^{2}]\Lambda(x_i);
\eea
\bea
\frac{1}{l_{mean}^{scatt,D}}\Big|_{LO}\simeq\f{1}{8\pi^{3}}n_{q_i}C_{F}G_{F}^{2}[(E_{\nu}
-\mu_ { \nu } )^ { 2 } +\pi^{2} T^{2}]\Lambda(x_i)g^{2}\text{log}\Big(\f{4g\mu}{\pi^{2}T}\Big);
\eea
\bea
\frac{1}{l_{mean}^{scatt,D}}\Big|_{NLO}&\simeq&\f{3}{2\pi}n_{q_i}C_{F}G_{F}^{2}[(E_{\nu}
-\mu_ { \nu } )^ { 2 } +\pi^{2} T^{2}]\Lambda(x_i)\Big[a_{1}g^{4/3}\Big(\f{T}{\mu}\Big)^{2/3}\nonumber\\
&+&a_{2}g^{2/3}\Big(\f{T}{\mu}\Big)^{4/3}+a_{3}\Big\{1-3\text{log}\Big(\frac{0.209g\mu}{T}\Big)\Big\}\Big(\f{T}{\mu}\Big)^{2}\Big]
\eea
where $n_{q_{i}}$ is the number density of quark of flavor i and $\sigma_0 \equiv 4G_{F}^2 m_{e}^2/\pi$ \cite{tubb75,lamb76}
 and $\Lambda(x_i)$ is defined in
\cite{pal11} where $x_i=\mu_{\nu}/\mu_{q_i}$ if $\mu_{\nu}< \mu_{q_i}$
and $x_i=\m_{q_i}/\mu_{\nu}$ if $\mu_{\nu}> \mu_{q_i}$.
$C_{v_i}$ and $C_{A_i}$ are the vector and axial vector coupling constants which are given in detail in Table (II) of \cite{Iwamoto:1982zz, Lattimer:1991ib, Tubbs:1975jx, Lamb:1976ac}.
%
The contributions from the Fermi liquid(FL), LO and NLO are added to obtain the MFP of the neutrinos.
\subsubsection*{The case of non-degenerate neutrinos}

We now derive MFP for non-degenerate neutrinos 
{\em i.e.} when $\mu_{\nu}\ll T$ beyond the Fermi-liquid contribution. For non-degenerate 
neutrinos the inverse process (\ref{inv}) is not considered as for un-trapped case, the reverse reaction is assumed to be zero. Hence, we neglect
the second term in the curly braces of Eq.(\ref{mfp01})\cite{Iwamoto:1982zz}. Including the strong interactions between the quarks results in a non vanishing squared matrix amplitude. The neutrino momentum in energy conserving relation due to the thermal production of the neutrinos can safely be neglected \cite{Iwamoto:1982zz}. The Fermi liquid case, LO and NLO are given as \cite{Iwamoto:1982zz, Lattimer:1991ib, pal11, adhya12, adhyaconf12},
\bea\label{mfp03}
\frac{1}{l_{mean}^{abs,ND}}\Big|_{FL}&=&\frac{3C_F\alpha_s}{\pi^4}G_{F}^2\cos^2\th_c
~\mu_d~\mu_u~\mu_e~\frac{(E_{\nu}^2+\pi^2 T^2)}{(1+e^{-\beta E_{\nu}})};
\eea
\bea
\frac{1}{l_{mean}^{abs,ND}}\Big|_{LO}\simeq\frac{C_{F}^{2}\alpha_s}{2\pi^6}G_{F}^2\cos^2\th_c\mu_{e}\frac{(E_{\nu}^2+\pi^2 T^2)}{(1+e^{-\beta E_{\nu}})}(g\mu)^{2}\text{log}\Big(\f{4g\mu}{\pi^{2}T}\Big);
\eea
\bea
\frac{1}{l_{mean}^{abs,ND}}\Big|_{NLO}&\simeq&\frac{3C_{F}^{2}\alpha_s}{\pi^4}G_{F}^2\cos^2\th_c\mu^{2}\mu_{e}\frac{(E_{\nu}^2+\pi^2 T^2)}{(1+e^{-\beta E_{\nu}})}\Big[b_{1}g^{4/3}\Big(\f{T}{\mu}\Big)^{2/3}\nonumber\\
&+&b_{2}g^{2/3}\Big(\f{T}{\mu}\Big)^{4/3}+b_{3}\Big\{1-3\text{log}\Big(\f{0.209g\mu}{T}\Big)\Big\}\Big(\f{T}{\mu}\Big)^{2}\Big]
\eea
where the constants are evaluated as $b_1=0.03, b_2=-0.149$ and $b_3=-0.073$ \cite{adhya12}. In the non-degenerate case, the FL term is proportional to $\alpha_s$ 
and is not same as the free Fermi gas. However, for the degenerate case,
the FL term contributes at the order $(\alpha_s)^0$.
The difference arises due to the evaluation of the squared matrix element for
the non-degenerate case where $v.p_e\sim C_F \alpha_s \mu_e/\pi$\cite{schafer04}.

Similarly, for the scattering of non-degenerate neutrinos in quark matter with appropriate phase space corrections we obtain \cite{Iwamoto:1982zz, Lattimer:1991ib, Tubbs:1975jx, Lamb:1976ac, pal11, adhya12, adhyaconf12},
\bea\label{mfp_scnd}
\frac{1}{l_{mean}^{scatt,ND}}\Big|_{FL}&=&\frac{C_{V_{i}}^2 +
C_{A_i}^2}{5\pi}n_{q_i}G_{F}^{2}\f{E_{\nu}^{3}}{\mu};
\eea
\bea
\frac{1}{l_{mean}^{scatt,ND}}\Big|_{LO}&\simeq&\frac{C_{V_{i}}^2 +
C_{A_i}^2}{30\pi^{3}}n_{q_i}G_{F}^{2}C_{F}\f{E_{\nu}^{3}}{\mu}g^{2}\text{log}\Big(\f{4g\mu}{\pi^{2}T}\Big);
\eea
\bea
\frac{1}{l_{mean}^{scatt,ND}}\Big|_{NLO}&\simeq&(C_{V_{i}}^2 + C_{A_i}^2)n_{q_i}G_{F}^{2}C_{F}\Big[b_{1}'\f{T^{2/3}g^{4/3}}{\mu^{5/3}}+b_{2}'\f{T^{4/3}g^{2/3}}{\mu^{7/3}}\nn\\
&&+b_{3}'\Big\{1-3\text{log}\Big(\f{0.209g\mu}{T}\Big)\Big\}\Big(\f{T^2}{\mu^3}\Big)\Big]
\eea
where $b_1'=0.002, b_2'=-0.009$ and $b_3'=-0.005$ \cite{adhya12}.
Here, it has been assumed $m_{q_i}/p_{f_i}\ll 1$.
Thus, the total MFP for non-degenerate neutrinos is obtained by summing up the contributions from the absorption and scattering parts to get the expression of the MFP of the non-degenerate neutrinos up to the NLO terms.
\subsection{Neutrino emissivity}
In this section, we focus on the calculation of the neutrino emissivity from the neutrino MFP. For our calculation, we consider the MFP of non-degenerate neutrinos.
The total emissivity of the non-degenerate neutrinos is obtained by multiplying the neutrino energy with the inverse of the MFP with appropriate factors and integrated over the neutrino momentum given by,
\bea
\veps=\int \frac{d^{3}p_{\nu}}{(2\pi)^{3}}E_{\nu}\frac{1}{l(-E_{\nu},T)}.
\eea
The total NFL correction can be obtained as,
\bea
\varepsilon - \varepsilon_{0} = \varepsilon_{LO} + \varepsilon_{NLO}
\eea
where,
\bea
\varepsilon_{0} \simeq \frac{457}{630}G_{F}^{2}cos^{2}\theta_{c}\alpha_{s}\mu_{e}T^{6}\mu^2
\eea
is the usual Fermi liquid contribution which agrees with the result presented in ref.\cite{Iwamoto:1982zz}.
At the LO we have obtained \cite{schafer04, adhya12, adhyaconf12},
\bea
\varepsilon_{LO} \simeq \frac{457}{3780}G_{F}^{2}cos^{2}\theta_{c}C_{F}\alpha_{s}\mu_{e}T^{6}\frac{(g\mu)^2}{\pi^2}\text{ln}\Big(\frac{4g\mu}{\pi^{2}T}\Big)
\eea
The NLO contribution to the neutrino emissivity is \cite{adhya12, adhyaconf12, Adhya:2015uaa},
\bea
\varepsilon_{NLO} &\simeq& \frac{457}{315}G_{F}^{2}cos^{2}\theta_{c}C_{F}\alpha_{s}\mu_{e}T^{6}\Big[c_{1}T^{2} \nn\\
&+& c_{2}T^{2/3}(g\mu)^{4/3} - c_{3}T^{4/3}(g\mu)^{2/3} - c_{4}T^{2}\text{ln}\Big(\frac{0.656g\mu}{\pi T}\Big)\Big]
\eea
where the constants are $c_1=-0.035, c_2=0.015, c_3=0.075$ and $c_4=-0.109$ \cite{adhya12}.
The NFL correction which appear in the phase space integral of the MFP \cite{pal11} and gradually in the expression of the emissivity \cite{schafer04} is actually related to the unscreened magnetic/ transverse interaction. The factor of $T^6$ can be understood as one power of $T$ is obtained from phase space integral of a degenerate fermion and $T^3$ from the phase space integral of the neutrino. One power of $T$ from the energy conserving $\delta$ function is canceled by a power from the emitted neutrino energy. For excitations lying close to the Fermi surface, the angular integrals provide no temperature dependence.
\subsubsection*{Emissivity of neutrinos from analogous processes}
In addition to the above mentioned quark direct process, there are contributions from other processes in different domain of density and momentum of the constituent particles. As for example, the neutrino emissivity from modified quark URCA process is given as \cite{Iwamoto:1982zz},
\bea
\varepsilon^{qURCA}\sim \alpha_c^2 G_F^2 \rm{cos}^2\theta_C p_F(q)(k_B T)^8
\eea
In lower densities, dominant neutrino emission mechanisms are the one-nucleon processes e.g., $n\to pe\bar{\nu}$, called nucleon direct URCA (DU) reactions. Their emissivity is\cite{Iwamoto:1982zz},
\bea
\varepsilon^{\rm DU}\sim 10^{27}\times T_9^6\,(n/n_0)^{2/3}\theta
(n -n_{c}^{\rm DU}) \,\rm erg/ \rm cm^3\, s
\eea
where $n$ is the nucleon density measured in  units of the nuclear matter saturation density
$n_0$. The DU processes are operative only when the proton fraction exceeds a critical value of
11--14\%. Another similar process of quite interest is the two nucleon interactions (in presence of a bystander particle) e.g., $nn\to
npe\bar{\nu}$, called modified Urca processes (MU) with the emissivity \cite{Iwamoto:1982zz},
\bea
\varepsilon^{\rm MU}\sim
10^{21}\times T_9^8\,(n/n_0)^{2/3} \,\rm erg/\rm cm^3\, s
\eea
However, other different processes, the nuclear bremstrahlung reactions, such as $nn\rightarrow nn\nu \bar{\nu}$ (nB), $np\rightarrow np\nu \bar{\nu}$ (npB) and $pp\rightarrow pp\nu \bar{\nu}$ (pB) have an order of magnitude smaller emission rates than the MU processes. 
In addition, quark matter at asymptotically high density is
in the color-flavor locked (CFL) phase~\cite{alford} of color
superconductivity (CSC). The CSC phases at moderate density are
still unclear due to the complicated non-perturbative effect.
Various candidates have been proposed, such as two flavor CSC
(2SC)~\cite{huang}, gapless 2SC~\cite{shovkovy}, gapless CFL
(gCFL)~\cite{alford1}, crystalline CSC~\cite{alford2}, and spin-1
CSC~\cite{iwasaki,schafer1,schmitt}. Neutrino emission from CSC
quark matter is exponentially suppressed at low temperature, if the
quasi-particle spectra are fully
gapped~\cite{alford3,jaikumar,schmitt2,anglani,wang,carter}. Similar
to nuclear matter, the pair breaking and recombination effect
results in an emissivity of $\epsilon\sim T^7$ at temperature close
to $T_c$~\cite{jaikumar2}. The emissivity behaves as $\epsilon\sim
T^{5.5}$ for gCFL~\cite{alford3} and $\epsilon\sim T^{6}$ for g2SC,
see, e.g. Table 1 of Ref.~\cite{schafer04}.
\subsection{Cooling behavior of neutron star with dense quark core}
\begin{figure}[]
\bigskip
\begin{center}
\includegraphics[height=5.5cm, angle=0]{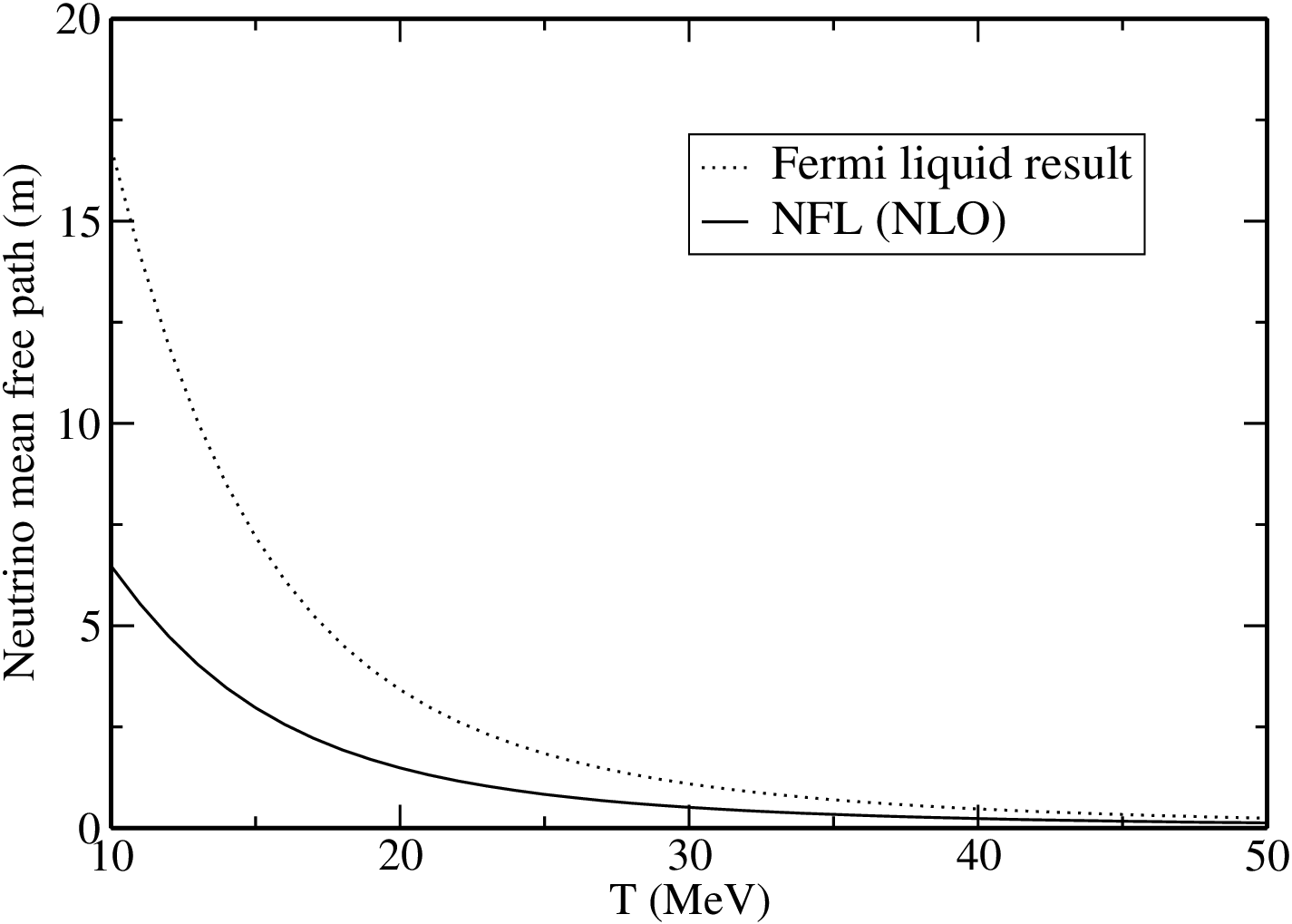}
~~~\includegraphics[height=5.5cm, angle=0]{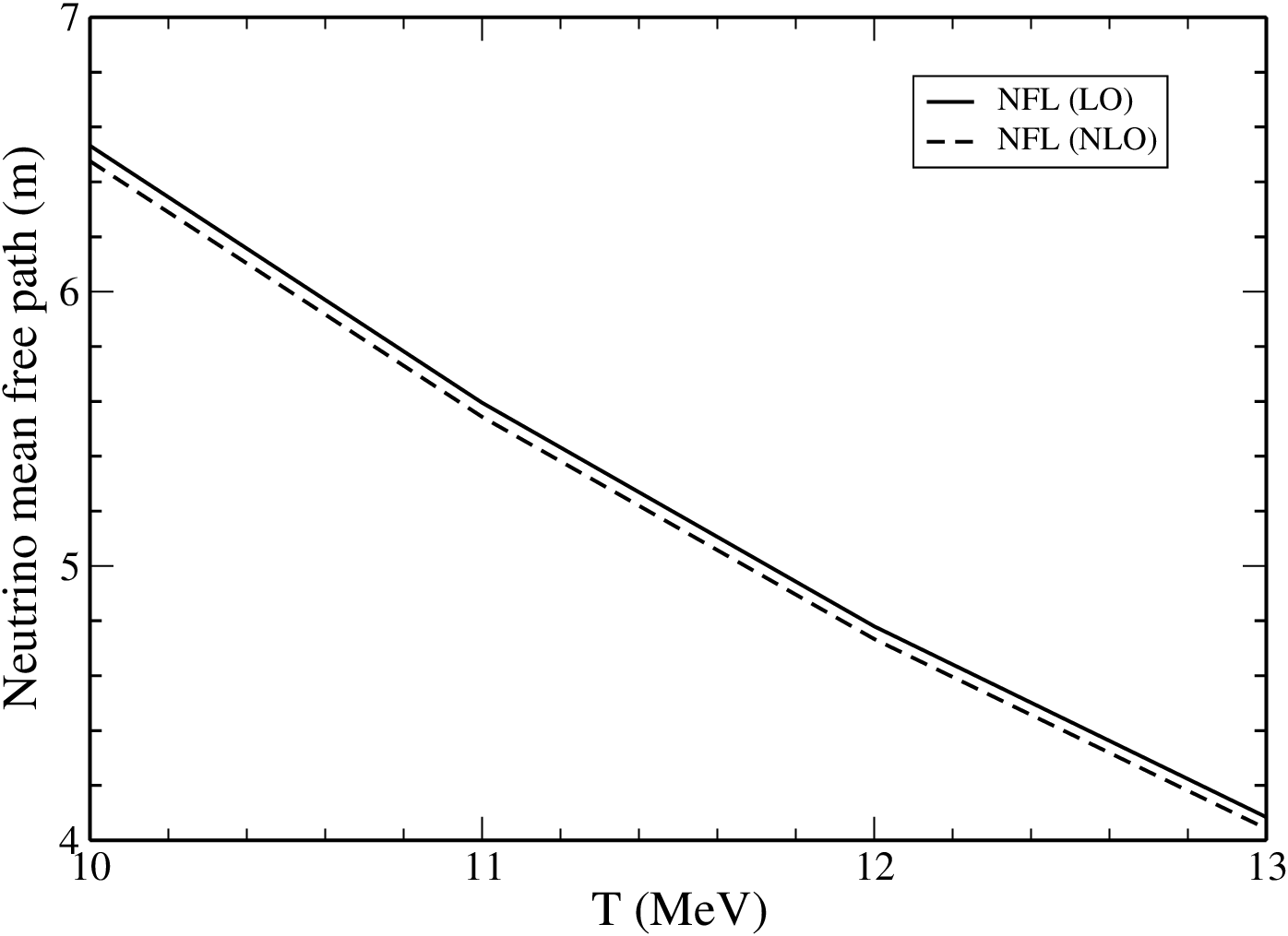}
\caption{\it{Mean free path of degenerate neutrinos. A comparison between the FL result and NLO corrections for the NFL effects has been shown in the left panel. The  reduction of the MFP due to NLO corrections has been shown in the right panel.}}
\label{figdn}
\end{center}
\end{figure}

\begin{figure}[]
\bigskip
\begin{center}
\includegraphics[height=5cm, angle=0]{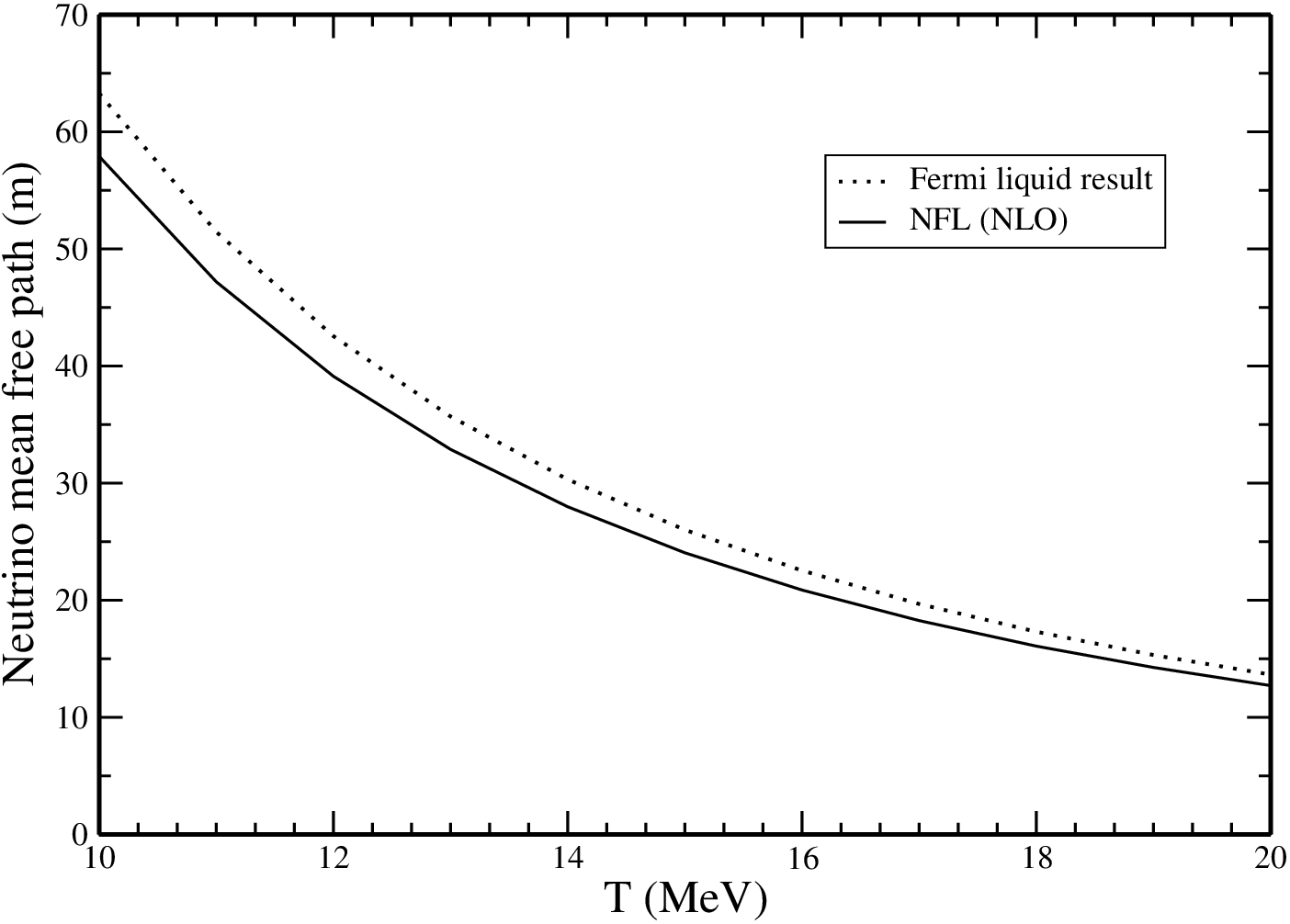}
~~~\includegraphics[height=5cm, angle=0]{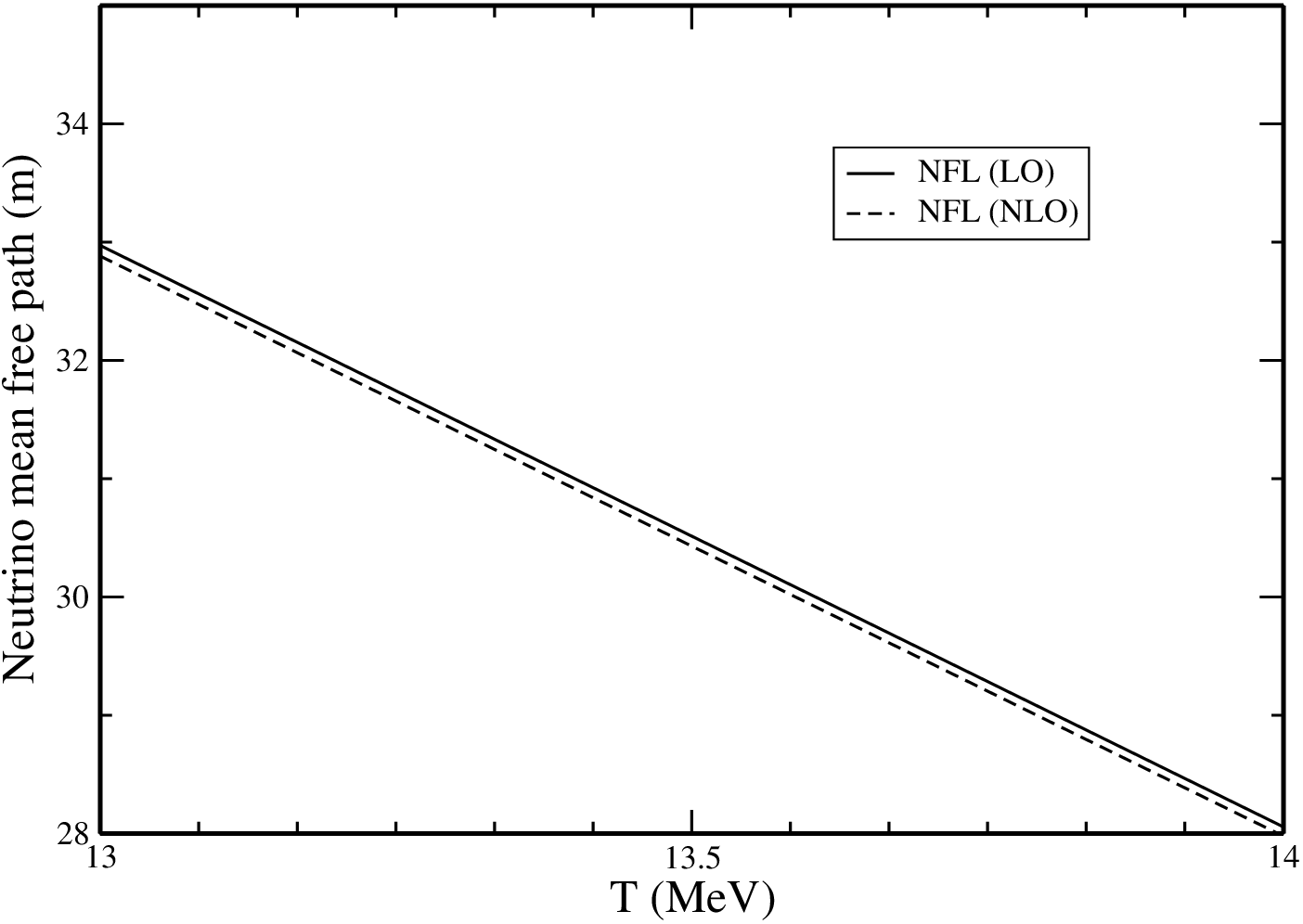}
\caption{\it{Mean free path of non-degenerate neutrinos. A comparison between the FL result and NLO corrections for the NFL effects has been presented in the left panel. The  reduction of the MFP due to NLO corrections has been shown in the right panel.}}
\label{figndn}
\end{center}
\end{figure}
\begin{figure}[htb]
\bigskip
\begin{center}
\includegraphics[height=5.5cm, angle=0]{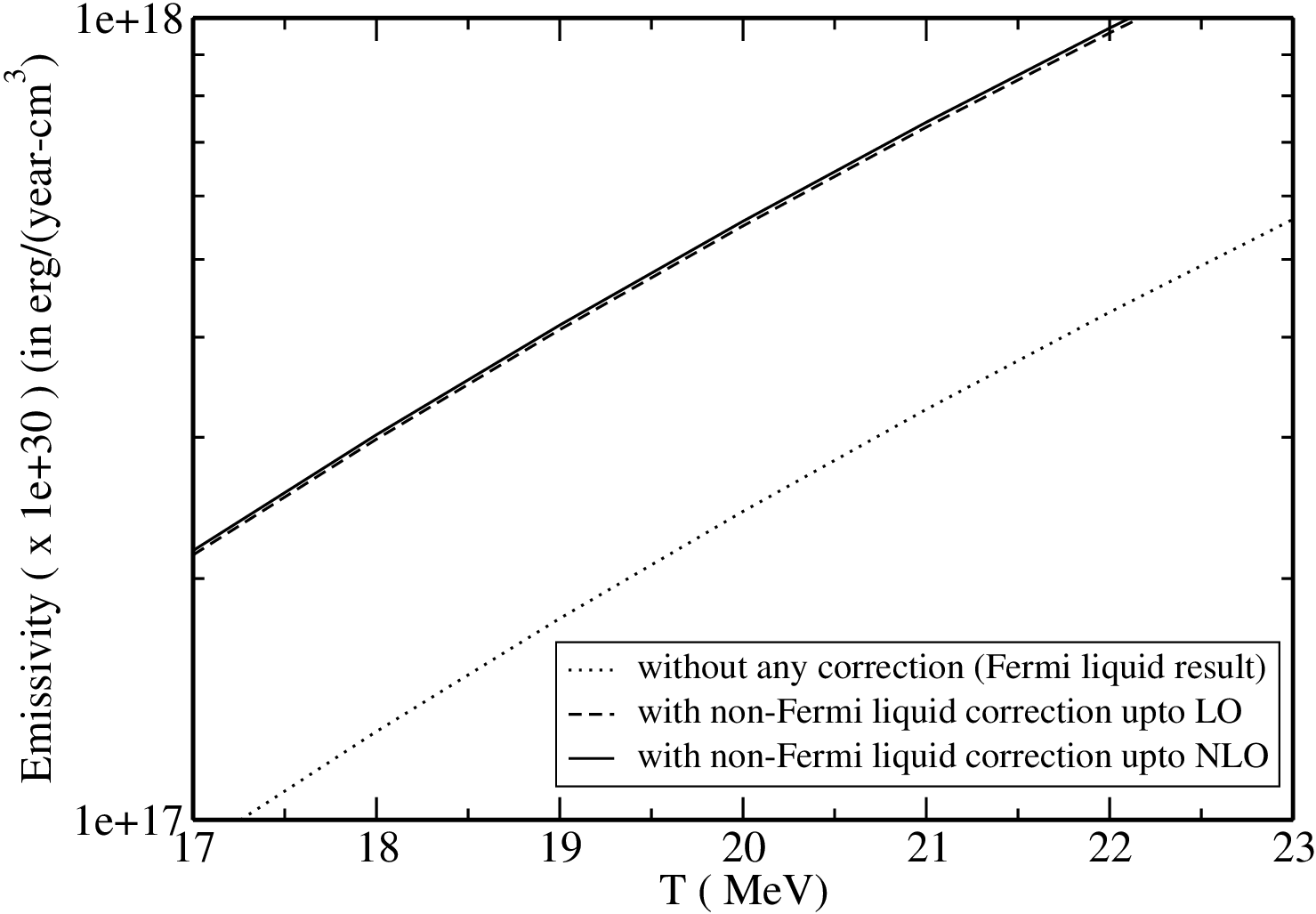}
~~~~~~~~~~\includegraphics[height=5.5cm, angle=0]{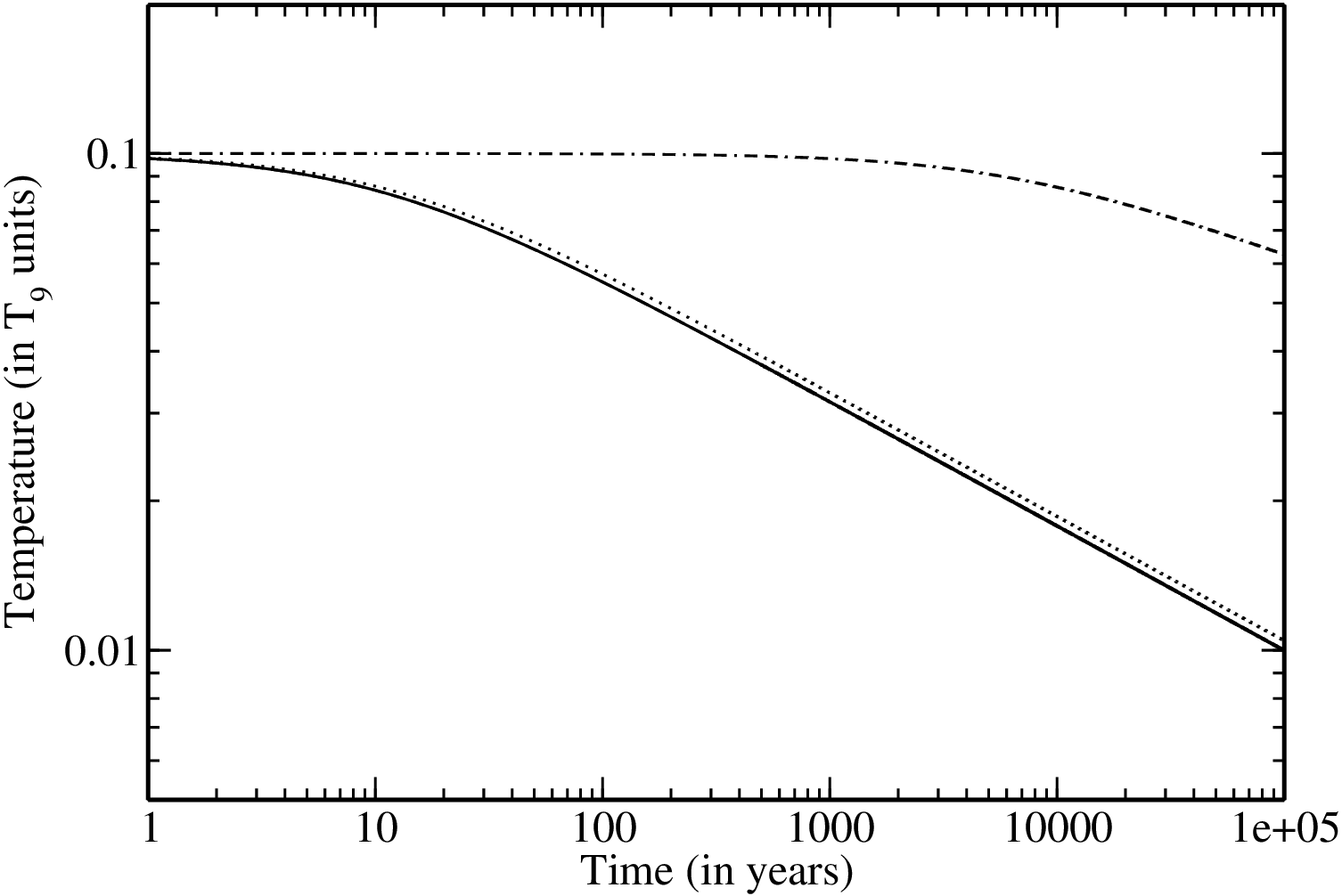}
\caption{\it{The  emissivity of the neutrinos with temperature in degenerate quark matter ($T_9$ in units of $10^9$ K) has been shown in the left panel. The cooling behavior of neutron star with core as neutron matter and degenerate quark matter has been presented in the right panel. The dotted line represents the FL result, 
the solid line depicts the NFL NLO correction. 
The dash-dotted line gives the cooling behavior of  
the neutron star core made up of purely neutron matter.}
\label{figecv}
\label{figcool}}
\end{center}
\end{figure}
The results of the MFP of neutrinos with the temperature has been presented in the Figs.(\ref{figdn}). A quark chemical potential of $500$ MeV is considered for the process. This agrees well with the high density $\sim 6\rho_{0}$ ($\rho_{0}$ is the nuclear matter saturation density) at the core.
We have  taken $\mu_{e}=15 MeV$ and $\alpha_{s}=0.1$. 
In left panel of Fig.(\ref{figdn}) we note that
there is a considerable decrease in MFP of degenerate neutrinos due to 
NLO corrections over the Fermi liquid result. 
In right panel of Fig.(\ref{figdn}) LO correction is compared with 
the NLO correction and it is seen that the MFP with NLO correction
is reduced marginally as compared to LO correction.
Similar features have been observed in case of the MFP of non-degenerate
neutrinos as displayed in Fig.(\ref{figndn}). This modest difference
in the MFP between NFL LO and NLO corrections also leads to marginal
difference in the emissivity for the two cases. This is shown in 
Fig.(\ref{figecv}). 
These small reductions are reflected in the marginally enhanced 
emissivity of the non-degenerate neutrinos which has been shown in 
left panel of Fig.(\ref{figecv}). We note that there is a modest increase 
in the emissivity of the neutrinos. 
The complicated cooling equation has to be solved numerically as shown in Fig.(\ref{figecv}) (right panel).
The temperature of the neutron star with a quark matter core shows a dependency with time. To analyze the cooling of the star \cite{peth04,finley92,rutledge01}, the specific heat capacity of the quark matter core needs to be taken into consideration along with the emissivity via the cooling process \cite{Iwamoto:1982zz, schafer04}. We have considered the NFL contribution to the specific heat capacity to compute our results.
We observe that the cooling of neutron star is marginally faster in case of NFL (NLO) as compared to the Fermi liquid result (shown in Fig.(\ref{figcool})). 
In addition, it is found that the cooling process from the quark core is much faster than the case where the nuclear matter is considered to be present at the core.
This influence of magnetic interaction has a significant effect on the cooling behavior of the neutron star, which in the present case, is  modified due to incorporation of the NFL contribution beyond LO. This leads to a marginally faster cooling of the neutron star due to neutrino emission URCA processes and the quark neutrino scattering reaction.

\subsection{Pulsar kick velocity through asymmetric neutrino emission}
The amount of pulsar acceleration depends on the polarization of the electron
spin and the momenta.
 The kick velocity can be written as \cite{sagert08, sagertarxiv1},
\begin{eqnarray}
 dv=\frac{\chi}{M_{NS}}\frac{4}{3}\pi R^{3}\varepsilon dt
 \label{diffv}
\end{eqnarray}
where the polarization fraction of the electrons has been denoted by $\chi$ and the neutrino emissivity by $\varepsilon$.
 Using the cooling equation, 
\begin{eqnarray}
 C_{v}^BdT=-\varepsilon dt,
 \label{cooleq}
\end{eqnarray}
One can rewrite Eq.(\ref{diffv}) in terms of the specific heat ($C_v$) of the quark matter core.
The influence of the exterior magnetic field on the specific heat of the quark matter in addition to the NFL effect can be explored in an interesting way. In the presence of a constant external magnetic field (B) along the z axis, the thermodynamic potential can be written as \cite{chakraborty96,bandopadhyay97},
\bea
\Omega^B=-\f{g_dT|q|B}{2\pi^2}\sum_{\nu=0}^{\infty}\int_0^\infty dp_z\log(1+e^{\beta(\mu-\epsilon)})
\eea
where $\epsilon=\sqrt{p_z^2+m^2+2\nu|q|B}$ is the single particle energy eigen value, $g_d$ is the quark degeneracy and $\nu=0,1,2,..$. 
Thus, the specific heat computed from the thermodynamic potential is \cite{adhya14,Adhya:2015uaa},
\bea
C_{v}\Big{|}_{FL}^{B}=\f{N_CN_fTm_q^2}{6}\Big(\f{B}{B_{cr}^q}\Big)
\label{cvFLB}
\eea
Including the effect of the NFL behavior in the specific heat capacity, through the in- medium quark dispersion relation,
\bea
\omega=(E_{p(\omega)} +{\rm Re}\Sigma(\omega,p(\omega)))
\eea
where the $\Sigma$ represents the one loop quark quasi-particle self energy,
the LO correction is given as \cite{adhya14},
\bea
C_v\Big{|}_{LO}^B\simeq\Big(\f{N_CN_fC_f\alpha_s}{36\pi}\Big)m_q^2\Big(\f{B}{B_{cr}^q}\Big)T\Big[(-1+2\gamma_E)+2log\Big(\f{2m_B}{T}\Big)\Big]
\eea
The NLO contribution to the specific heat capacity is obtained as,
\bea
C_v\Big{|}_{NLO}^B\simeq&&\Big(\f{N_CN_f}{3}\Big)(C_f\alpha_s)\Big(m_q^2\f{B}{B_{cr}^q}\Big)T\Big[c_1\Big(\f{T}{m_B}\Big)^{2/3}\nn\\
&+&c_2\Big(\f{T}{m_B}\Big)^{4/3}+c_{3}\Big(\f{T}{m_B}\Big)^{2}\Big(c_{4}-\log\Big(\f{T}{m_B}\Big)\Big)\Big]
\eea
The in-medium propagators contain the dressed mass  $m_B$ (Debye mass in the QCD case) in presence of magnetic field,
\bea
m_B^{2}=\f{N_fg^2m_q^2}{4\pi^2}\Big(\f{B}{B_{cr}^q}\Big)
\eea
The kick velocity taking into account the magnetic field effect on the specific heat capacity of the quarks reads as \cite{adhya14},
\bea
 v\Big|_{FL}^B &\simeq&\frac{4.15 N_{C}N_f}{3}\Big(\frac{\sqrt{m_q^2(B/B_{cr}^q)}}{400MeV}\frac{T}{1MeV}\Big)^{2}\Big(\frac{R}{10km}\Big)^{3}\frac{1.4M_{\odot}}{M_{NS}}\chi\frac{
km}{s}
\label{vBFL}
\eea
Incorporation of the anomalous effect, the LO contribution to the kick velocity is\cite{adhya14,Adhya:2015uaa},
\bea
v\Big|_{LO}^B &\simeq&\frac{8.8 N_{C}N_f}{3}(C_f\alpha_s)\Big(\frac{\sqrt{m_q^2(B/B_{cr}^q)}}{400MeV}\frac{T}{1MeV}\Big)^{2}\Big(\frac{R}{10km}\Big)^{3}\nn\\
&&\times\frac{1.4M_{\odot}}{M_{NS}}\Big[0.0635+0.05\log\Big(\f{m_B}{T}\Big)\Big]\chi\frac{
km}{s}
\label{vBLO}
\eea
The calculation has been extended beyond the LO in NFL correction.
The NLO correction is \cite{adhya14,Adhya:2015uaa},
\bea
v\Big|_{NLO}^B&\simeq&\frac{8.3 N_{C}N_{f}}{3}\Big(\f{B}{B_{cr}^q}\Big)\Big(\frac{m_{q}}{400MeV}\frac{T}{1MeV}\Big)^{2}\Big(\frac{R}{10km}\Big)^{3}\frac{1.4M_{\odot}}{M_{NS}}\nn\\
&&\times\chi(C_{F}\alpha_{s})\Big[a_1\Big(\frac{T}{m_B}
\Big)^{2/3}+a_2\Big(\frac{T}{m_B}\Big)^{4/3}\nn\\
&&+ \Big[a_3+a_4\ln\Big(\frac
{m_B}{T}\Big)\Big]\Big(\frac{T}{m_B}\Big)^2\Big]\frac{km}{s}
\label{vBNLO}
\eea
The long range magnetic interactions are responsible for the anomalous $T^2\rm{ln}T^{-1}$ term in the expressions of the pulsar kick velocity.

\begin{figure}[]
\bigskip
 \bigskip
\begin{center}
\resizebox{5.0cm}{5.0cm}{\includegraphics{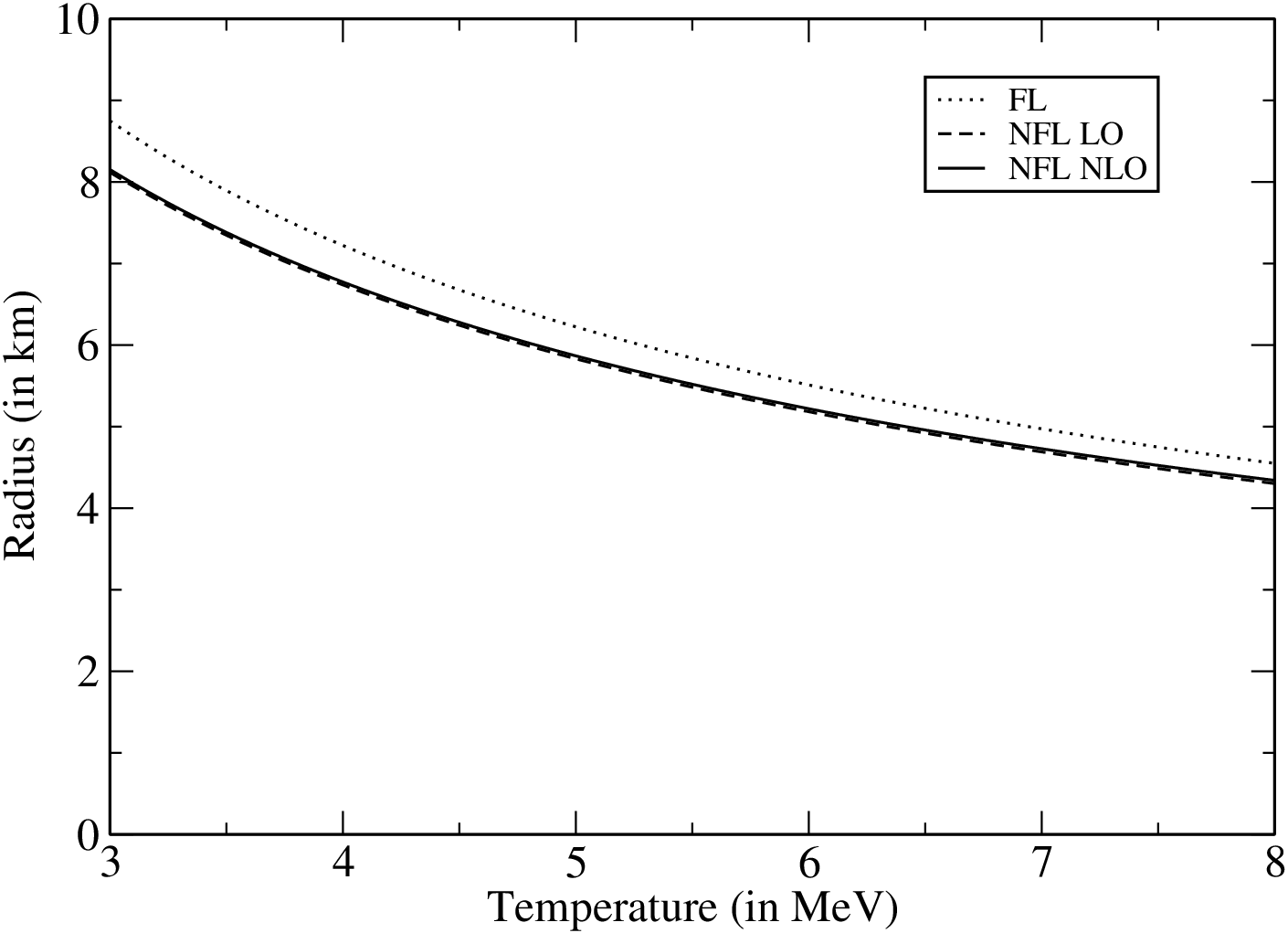}}~~~~~~~\resizebox{5.0cm}{5.0cm}{\includegraphics{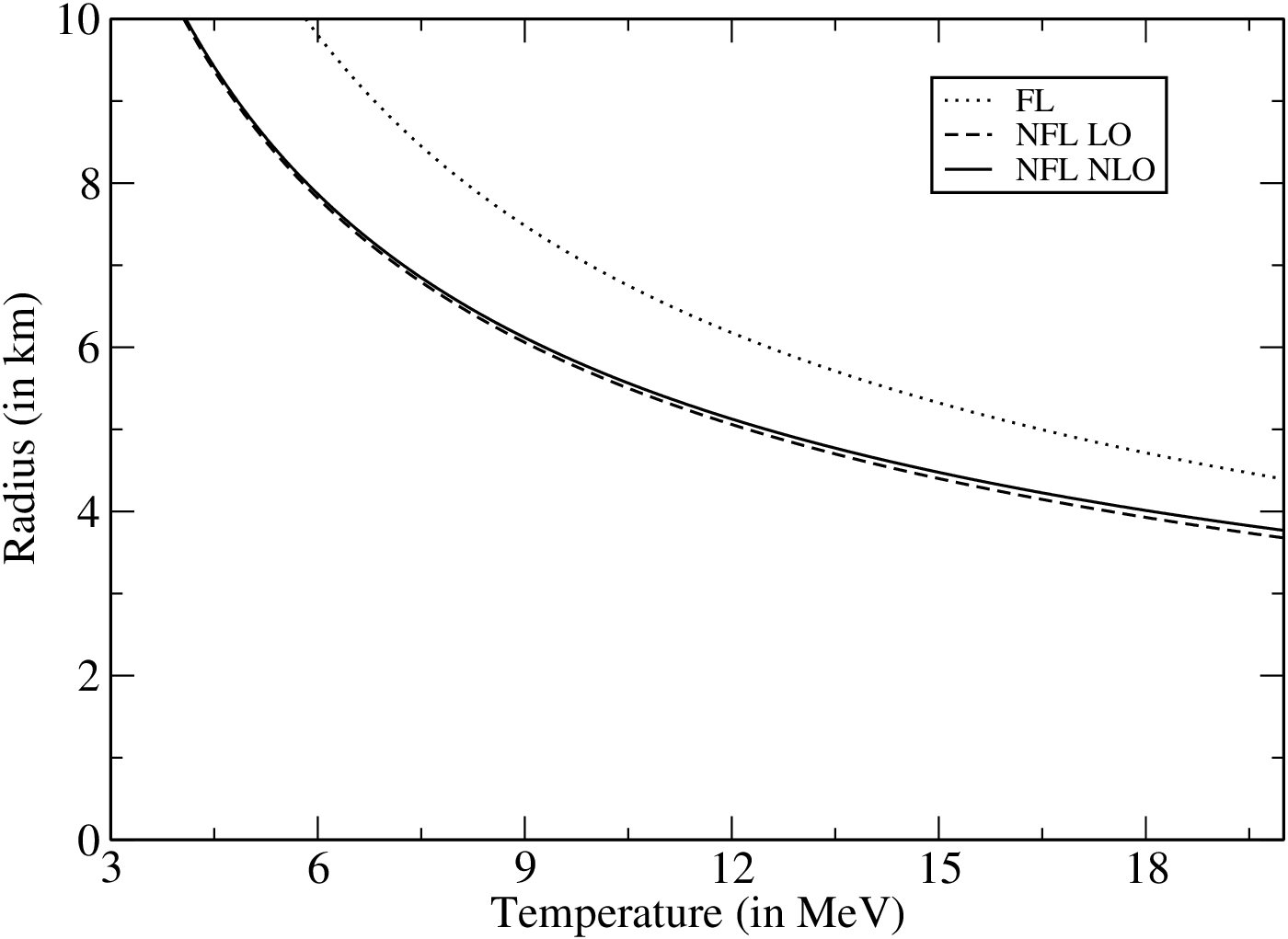}}~~~~~~~\resizebox{5.0cm}{5.0cm}{\includegraphics{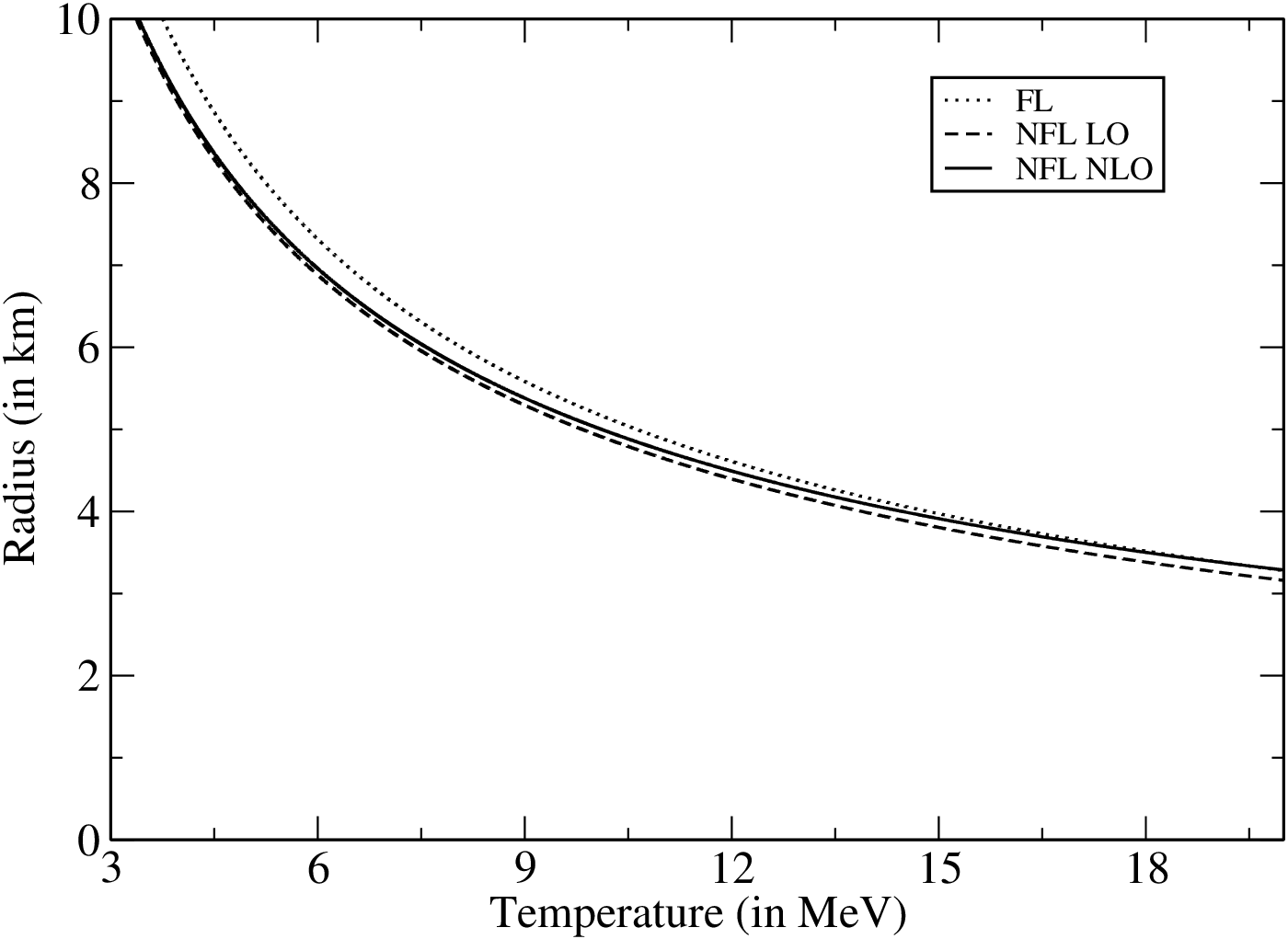}}
\caption{\small{The comparison of the relationship (FL,LO and NLO respectively)  where high magnetic field $(B=10^{19} G)$ has been taken into account along with vanishing temperature for kick velocity of $100 km/s$  has been shown in the extreme left panel. The middle panel shows the corresponding case when high magnetic field effect in specific heat is included.
The extreme right figure shows the comparison between the FL, NFL LO and NFL NLO result for the radius and temperature dependence for partially polarized electrons in presence of weak magnetic field $(B=5\times10^{15} G)$ for kick velocity of $100 km/s$. }}
\label{figkickvel}
\end{center}
\end{figure}

For the case of cold neutron stars, the spin polarization of electrons is given by \cite{sagertarxiv1},
\cite{sagert08},
\begin{eqnarray}
 \chi\simeq\frac{3}{2}\frac{m_{e}^2}{\mu_{e}^2-m_{e}^2}\Big(\frac{B}{B_{cr}^e}
\Big)
\end{eqnarray}
where the critical value of the magnetic field is taken to be
$B_{cr}^e\simeq4.4\times10^{13} G$ to Landau quantize the electrons. The weak magnetic field refers to $(\mu_e^2-m_e^2)>(2eB)$ for partially occupied Landau levels which makes the spin polarization fraction $\chi<<1$.
The second case arises when the strength of the magnetic field is chosen to be much larger than the
temperature, the chemical potential and the electron mass ($\mu_e, m_e,T
\ll\sqrt{2eB}$) for $\chi$ close to unity. The electron polarization is given as \cite{sagert08},
\begin{eqnarray}
\chi&\sim& 1-\frac{4}{\ln(2)}\sqrt{\frac{\pi T}{2\sqrt{2eB}}}e^{-\sqrt{2eB}/T}.
\label{pol_4}
\end{eqnarray}
For the case of large magnetic field, the kick velocity cannot be solved analytically.
The net contribution to the pulsar kick velocity up to NLO is obtained by the sum of the
Fermi liquid result and the non-Fermi liquid correction up to NLO:
\bea
v\Big{|}_{total}^B=v\Big|_{FL}^B+v\Big|_{LO}^B+v\Big|_{NLO}^B
\label{vtotal}
\eea
With these results, an estimation of the quark phase radius of the NS with the temperature of the quark matter in the core is presented. 
In Fig.(\ref{figkickvel}), for the case of highly polarized electrons,  a comparison between FL, LO and NLO corrections to the kick velocity is shown in the extreme left panel. The middle panel shows an exactly similar behavior with inclusion of the magnetic field on the specific heat capacity of quarks.
In the figure at extreme right, similar behavior for NFL cases are observed when the electrons are partially polarized for low magnetic fields.

  It is observed that the pulsar kick velocity receives significant contribution from the logarithmic corrections at the LO in comparison to the FL result. The results are computed with LO and NLO corrections to include plasma or quasi-particle effects which are anomalous (NFL) effects. In addition, comparison has been made between the NFL LO and NLO contributions to the kick velocity with the FL case. It is seen that the NFL LO corrections are significant while calculating the radius-temperature relationship as seen in Fig.(\ref{figkickvel}). The anomalous corrections introduced to the pulsar kick velocity due to the NFL (LO) behavior is responsible for the appreciable increment of the kick velocity for a particular value of temperature found in the NS core. However, for all the cases, moderate change in the R-T relationship has been observed for the NLO correction with respect to the LO case.
\section*{Conclusions}
 \noindent
 The prime objective of this present review is to present a comprehensive description of the collective excitations involving quasi-particles in the domain of relativistic dense plasma. It will not be out of context to recall that such ultra-degenerate system is relevant for the astrophysical scenario of the QCD phase diagram \rm{i.e.} neutron stars. Thus, different physical quantities of interest have been studied for such system involving the quark direct URCA and quark- neutrino scattering reactions. We have found the MFP for both the degenerate as well as the non-degenerate neutrinos involving terms which have fractional powers in $(T / \mu )$ at higher orders using the perturbation theory and the NFL phenomenon as discussed above. In
addition, we have also calculated the emissivity of neutrinos
and examined NLO corrections over NFL LO and the
simple FL case. Finally, we have examined the cooling
behavior of the neutron star involving NLO correction
to the emissivity of neutrinos as well as specific heat capacity of quark core. It might be noted that even though there is
a modest correction to the quantities like MFP and 
emissivity of neutrinos over LO and FL case, there is a
marginal alteration in the cooling behavior due to such
NFL corrections. Although the NLO corrections are not quantitatively significant, these corrections
at the higher order terms of the low temperature expansion involves
fractional powers of $T$ as well as subsequent logarithmically enhanced term $T^3ln(1/T)$
which are important from the theoretical aspect of NFL formalism for temperature tending to zero. The dynamical screening along with Pauli blocking is a distinctive feature of cold and dense plasma unlike the case of high temperature.

The extension of the NFL phenomena to the study of pulsar kick velocities show that it receives significant contribution from the anomalous leading logarithmic correction. Further, incorporation of the modified quark dispersion relations up to the next-to-leading logarithmic order to the kick velocity has been studied. In addition, we have analyzed the effect of external magnetic field on the specific heat capacity of quarks which modifies the pulsar kick velocity. The primary mechanism for the generation of kick velocity is asymmetric neutrino emission. This asymmetry can be attributed to the strong magnetic field which can align the electron spin
opposite to the magnetic field direction. Thus, we have taken into account the contribution from electron polarization fractions for different field strengths. We note that the presence of the logarithmic term and the fractional powers carrying signature of the weaker dynamical screening along with external magnetic field considerably enhances the kick velocity of the neutron star.

These results provide a novel initiation for the study of the interplay of strong magnetic fields \cite{adhya16} with dense QCD matter which can have significant impact in the context of interesting astrophysical entities such as neutron star.
\\
\\
\textit{Acknowledgments:}~\small{[SPA] would like to thank Prof. (Late) Abhee K. Dutt-Mazumder and Prof P. K. Roy without whose supervision this work would not have been possible. [SPA] would like to thank Prof. T. K. Nayak for continuous academic suppport during this work. The author  [SPA] declares that there is no conflict of interest regarding the publication of this paper. }


\section*{References}

\begin{thebibliography}{50}
\bibitem{shapiro_book} S.L.Shapiro and S.A.Teukolsky, {\it Black Holes, 
 White Dwarfs and Neutron Stars.} Wiley-Interscience, New York (1983).

\bibitem{MSTV90} A.B.~Migdal, E.E.~Saperstein, M.A.~Troitsky, and D.N.~Voskresensky, Phys. Rept. {\bf 192}, 179 (1990).

\bibitem{YLS99}D.G.~Yakovlev, A.D.~Kaminker, O.Y.~Gnedin, and P.~Haensel, Phys. Rept. {\bf 354}, 1
(2001).

\bibitem{V01} D.N.~Voskresensky, Lect. Notes Phys. {\bf 578},  467 (2001).

\bibitem{PGW} D.~Page, U.~Geppert, and F.~Weber, Nucl. Phys. A {\bf 777}, 497 (2006).
\bibitem{Sedr07} A.~Sedrakian, Prog. Part. Nucl. Phys. {\bf 58}, 168 (2007).
\bibitem{Huang:2007jw} 
  X.~g.~Huang, Q.~Wang and P.~f.~Zhuang,
  Phys.\ Rev.\ D {\bf 76}, 094008 (2007)


\bibitem{witten84} E.~Witten, Phys. Rev. D {\bf 30}, 272 (1984).
\bibitem{farhi84} E.~Farhi and R. L.~Jaffe, Phys. Rev. D {\bf 30}, 2379 (1984).



\bibitem{Gan:1993}
J. Gan and E. Wong, Phys. Rev. Lett. {\bf 71},  4226  (1993).

 


\bibitem{Chakravarty:1995}
S. Chakravarty, R.~E. Norton, and O.~F. Sylju{\aa}sen, Phys. Rev. Lett. {\bf
  74},  1423  (1995).

\bibitem{Reizer:1989}
M.~Y. Reizer, Phys. Rev. {\bf B40},  11571  (1989).


\bibitem{Varma:1989}
C.~M. Varma {\it et~al.}, Phys. Rev. Lett. {\bf 63},  1996  (1989).

\bibitem{Polchinski:1992ed}
J. Polchinski, Effective field theory and the {F}ermi surface, hep-th/9210046.

\bibitem{Polchinski:1994ii}
J. Polchinski, Nucl. Phys. {\bf B422},  617  (1994).

\bibitem{Nayak:1994ng}
C. Nayak and F. Wilczek, Nucl. Phys. {\bf B430},  534  (1994).
\bibitem{manuel00} C.Manuel, Phys.Rev.D {\bf 62}, 076009 (2000).
\bibitem{Bellac97} M. Le Bellac and C. Manuel, Phys. Rev. D {\bf 55}, 3215(1997).



\bibitem{schafer04} Sch\"{a}fer T and Schwenzer K 2004 {\it Phys.Rev.D} {\bf 70}
114037.
\bibitem{pal11} Pal K and Dutt-Mazumder A K 2011 {\it Phys.Rev.D} {\bf 84} 034004.
\bibitem{rebhan05} Gerhold A and Rebhan A 2005 {\it Phys.Rev.D} {\bf 71} 085010.
\bibitem{holstein73} Holstein T, Norton R E and Pincus P 1973 {\it Phys. Rev. B} {\bf
8} 2649.
\bibitem{hieselberg93} Hieselberg H and Pethick C J 1993 {\it Phys. Rev. D} {\bf 48}, 
2916.
\bibitem{sarkar10} Sarkar S and Dutt-Mazumder A K 2010 {\it Phys.Rev.D} {\bf 82}
056003.
\bibitem{sarkar11} Sarkar S and Dutt-Mazumder A K 2011 {\it Phys.Rev.D} {\bf 84}
096009.
\bibitem{sarkar13} Sarkar S and Dutt-Mazumder A K 2013 {\it arXiv 1209.5153}.
\bibitem{ipp04} Gerhold A, Ipp A and Rebhan A 2004 {\it Phys.Rev.D} {\bf 70} 105015
; {\bf 69} R011901.
\bibitem{sagert08} Sagert I and Schaffner-Bielich J 2008 {\it J. Phys. G} {\bf 35} 014062.
\bibitem{sagertarxiv1} Sagert I and Schaffner-Bielich J {\it arXiv 0708.2352}.


\bibitem{Yakovlev:2004iq} 
  D.~G.~Yakovlev and C.~J.~Pethick,
  Ann.\ Rev.\ Astron.\ Astrophys.\  {\bf 42}, 169 (2004).

\bibitem{finley92} J.P.Finley et al, The astrophysical Journal
{\bf 394}, L21 (1992).

\bibitem{Anglani:2006br}
  R.~Anglani, G.~Nardulli, M.~Ruggieri and M.~Mannarelli,
  Phys.\ Rev.\ D {\bf 74} (2006) 074005.

\bibitem{adhya12} Adhya S P, Roy P K and Dutt-Mazumder A K 2012 {\it Phys.Rev.D} {\bf 86} 034012.
\bibitem{adhya14} Adhya S P, Roy P K and Dutt-Mazumder A K 2014 , {\it J. Phys. G} {\bf 41}, 
025201.
\bibitem{Adhya:2015uaa} 
  S.~P.~Adhya and P.~K.~Roy,
  EPJ Web Conf.\  {\bf 95}, 04001 (2015).

\bibitem{Iwamoto:1982zz}
  N.~Iwamoto,
  Annals Phys.\  {\bf 141} (1982) 1.

\bibitem{Lattimer:1991ib}
  J.~M.~Lattimer, M.~Prakash, C.~J.~Pethick and P.~Haensel,
  Phys.\ Rev.\ Lett.\  {\bf 66} (1991) 2701.

\bibitem{Tubbs:1975jx}
  D.~L.~Tubbs and D.~N.~Schramm,
  Astrophys.\ J.\  {\bf 201} (1975) 467.

\bibitem{Lamb:1976ac} 
  D.~Q.~Lamb and C.~J.~Pethick,
  Astrophys.\ J.\  {\bf 209}, L77 (1976).

\bibitem{tatsumi09} K.Sato and T.Tatsumi, Nucl.Phys.A {\bf 826}, 74 (2009).
\bibitem{adhyaconf12} Adhya S P, Roy P K and Dutt-Mazumder A K 2013, {\it AIP Conf. Proc.} {\bf 1524}, pp. 263-266.


\bibitem{tubb75} D.L.Tubbs and D.N.Schramm, Astrophys.J.{\bf 201}, 467 (1975).
\bibitem{lamb76} D.Q.Lamb and C.J.Pethick, Astrophys.J.Lett.{\bf 209}, L77 (1976).

\bibitem{alford}M. Alford, K. Rajagopal and F. Wilczek,
NPB{\bf537} 443 (1999).

\bibitem{huang}M. Huang, IJMPE{\bf14}675 (2005).

\bibitem{shovkovy}I. A. Shovkovy and M. Huang, PLB{\bf564} 205 (2003); M. Huang and I. A. Shovkovy, NPA{\bf729} 835(2003).

\bibitem{alford1}M. Alford, C. Kouvaris and K. Rajagopal,
PRL{\bf92} 222001 (2004).

\bibitem{alford2}M. Alford, J. Bowers and K. Rajagopal,
PRD {\bf63} 074016 (2001).


\bibitem{iwasaki}M. Iwasaki and T. Iwado, PLB{\bf350} 163 (1995);
R. D. Pisarski and D. H. Rischke, PRD
{\bf61} 074017 (2000); M. G.
Alford, $et.\; al.$, PRD {\bf67} 054018 (2003).

\bibitem{schafer1}T. Sch\"afer, PRD {\bf62} 094007 (2000).

\bibitem{schmitt}A. Schmitt, Q. Wang and D. H. Rischke,
PRD{\bf66} 114010 (2002); PRL {\bf91} 242301 (2003); A. Schmitt,
PRD{\bf71} 054016 (2005).



\bibitem{alford3}M. Alford, $et.\; al.$,
PRD {\bf71} 114011 (2005).

\bibitem{schmitt2}A. Schmitt, I. A. Shovkovy and Q. Wang,
PRD{\bf73} 034012 (2006).

\bibitem{jaikumar}P. Jaikumar, C. D. Roberts and A. Sedrakian,
PRC{\bf73} 042801 (2006).

\bibitem{anglani}R. Anglani, $et.\; al.$,
PRD {\bf74} 074005 (2006).

\bibitem{wang}Q. Wang, Z. G. Wang and J. Wu,
PRD{\bf74} 014021 (2006).

\bibitem{carter}G. W. Carter and S. Reddy, PRD{\bf62} 103002 (2000).


\bibitem{jaikumar2}P. Jaikumar and M. Prakash, PLB{\bf516} 345 (2001).
\bibitem{peth04} D.G.Yakovlev and C.J.Pethick, Ann.Rev.Astron.Astrophys. 
{\bf 42} 169 (2004).
\bibitem{rutledge01} R.E.Rutledge et al, The astrophysical Journal {\bf 551}, 
921 (2001).



\bibitem{dorofeev85} Dorofeev O F, Radionov V N and Ternov I M 1985 {\it Soviet
Astronomy Letters} {\bf 11} 123.


\bibitem{chakraborty96} Chakraborty S 1996 {\it Phys.Rev.D} {\bf 54} 2.
\bibitem{bandopadhyay97} Bandopadhyay D, Chakraborty S and Pal S 1997 {\it Phys. Rev. Lett.} {\bf 79} 12.

\bibitem{adhya16} Adhya S P, Mandal M, Biswas S and Roy P K 2016 {\it Phys. Rev. D} {\bf 93}, 
074003.








\end{thebibliography}
\end{document}